\documentclass[aps,prl,twocolumn,english]{revtex4-1}

\usepackage{latexsym}
\usepackage{ifpdf}
\usepackage{graphicx}
\usepackage{epsfig}
\usepackage{amsmath, amssymb, bbm, mathrsfs}
\usepackage{multirow}
\usepackage{ulem}
\usepackage{fancyhdr}
\usepackage{color}
\usepackage{bm}
\usepackage{eurosym}
\usepackage{subfigure}
\usepackage{comment}
\usepackage{scrhack}
\usepackage{physics}
\usepackage{balance}
\usepackage{float}
\usepackage{flushend}
\usepackage{setspace}   

\usepackage{dcolumn}    
\usepackage{commath}   
\usepackage{subfigure}    
\usepackage{braket}
\usepackage{siunitx}
\usepackage{hyphenat}
\usepackage{array}  
\newcolumntype{C}[1]{>{\centering\arraybackslash\hspace{0pt}}p{#1}}

\usepackage[unicode=true,pdfusetitle,
 bookmarks=false,
 breaklinks=true,pdfborder={0 0 0},pdfborderstyle={},backref=false,colorlinks=true]
 {hyperref}
\hypersetup{
 colorlinks,linkcolor=red,citecolor=blue}

\newcommand{\CMT}[1]{{}}

\newcommand{\beginsup}{%
        \setcounter{table}{0}
        \renewcommand{\thetable}{S\Roman{table}}%
        \setcounter{figure}{0}
        \renewcommand{\thefigure}{S\arabic{figure}}%
        \setcounter{equation}{0}
        \renewcommand{\theequation}{S\arabic{equation}}%
        \setcounter{enumiv}{0}
        \renewcommand{\theenumiv}{S\arabic{enumiv}}%
        \setcounter{secnumdepth}{1}
}

\begin{document}


\title{Superconducting cavity piezo-electromechanics: the realization of an acoustic frequency comb at  microwave frequencies}

\author{Xu Han}
\thanks{Present address: Center for Nanoscale Materials, Argonne National Laboratory, Argonne, Illinois 60439, USA}
\affiliation{Department of Electrical Engineering, Yale University, New Haven, Connecticut 06520, USA}

\author{Chang-Ling Zou}
\affiliation{Department of Electrical Engineering, Yale University, New Haven, Connecticut 06520, USA}

\author{Wei Fu}
\affiliation{Department of Electrical Engineering, Yale University, New Haven, Connecticut 06520, USA}

\author{Mingrui Xu}
\affiliation{Department of Electrical Engineering, Yale University, New Haven, Connecticut 06520, USA}

\author{Yuntao Xu}
\affiliation{Department of Electrical Engineering, Yale University, New Haven, Connecticut 06520, USA}

\author{Hong X. Tang}
\email{hong.tang@yale.edu}
\affiliation{Department of Electrical Engineering, Yale University, New Haven, Connecticut 06520, USA}


\begin{abstract}
We present a nonlinear multimode superconducting electroacoustic system, where the interplay between superconducting kinetic inductance and piezoelectric strong coupling establishes an effective Kerr nonlinearity among multiple acoustic modes at 10 GHz that could hardly be achieved via intrinsic mechanical nonlinearity. By exciting this multimode Kerr system with a single microwave tone, we further demonstrate a coherent electroacoustic frequency comb and provide theoretical understanding of multimode nonlinear interaction in the  superstrong coupling limit. This nonlinear superconducting electroacoustic system sheds light on the active control of multimode resonator systems and offers an enabling platform for the dynamic study of microcombs at microwave frequencies.

\end{abstract}


\maketitle

\textit{Introduction.--} Mechanical and acoustic resonators with low dissipation have become pivotal elements in many applications such as weak mass and force detection \cite{Abbott2017,Goryachev2014,Jensen2008,Moser2013,Krause2012,Fu2019}, timing and frequency control \cite{Nguyen2007,Beek2012}, and frequency conversion \cite{Hill2012,Andrews2014,Higginbotham2018,Forsch2019,Jiang2020,Han2020,Han2021}. Recent investigations of the nonlinear behaviors in mechanical resonators \cite{Lifshitz2009,Rhoads2010,Noel2017}, especially at nano-/micro-scales, have attracted great attentions for their potential in novel device functionality; by exploiting the mechanical nonlinearity, parametric amplification \cite{Aldridge2005}, frequency tuning \cite{Gajo2020}, and frequency stabilization \cite{Antonio2012} have been realized.

For many applications, high operation frequencies above gigahertz \cite{Eichenfield2009,Han2014,Han2015} are highly desired for suppressing the impact of thermomechanical noise, improving the time-keeping precision, and preparing non-classical mechanical states.
However, so far, most nonlinear mechanical or acoustic systems can only operate at low frequencies up to a few megahertz.
The challenges arise from the preparation of high-quality (\textit{Q}) factor mechanical resonators at such high frequencies, as the \textit{Q} factor quickly reduces with frequency, limited by the empirical $f\cdot Q$ product \cite{Chandorkar2008}. At the same time, the intrinsic mechanical nonlinearity also degrades with increasing frequency, since the mechanical resonators become stiffer with their deformation amplitudes significantly reduced. 
As a result, the intrinsic nonlinear effects of gigahertz mechanical resonators, such as the commonly exploited Duffing \cite{Lifshitz2009} and electrostatic spring effects \cite{Unterreithmeier2009,Poot2015}, are typically very weak. To overcome these obstacles, it is of great importance to develop a hybrid mechanical system where nonlinearity can be substantially enhanced.

In this work, we demonstrate such a nonlinear superconducting electroacoustic system at 10 GHz where strong nonlinearity is achieved in the multimode strong coupling regime \cite{Han2016}. By engineering the kinetic inductance of a superconducting resonator, we establish effective Kerr nonlinear interactions \cite{Yurke2006} among multiple high-\textit{Q} acoustic modes in a bulk acoustic resonator (BAR). Leveraging this nonlinear multimode system, we demonstrate a coherent electroacoustic frequency comb driven by a single microwave parametric pump. We further study the coherence of the frequency comb by characterizing the stable relative phase of each comb line via time-domain measurement.

Phononic frequency combs have been recently investigated in micro-electromechanical systems (MEMS). These MEMS combs operate at megahertz frequencies in order to harness the intrinsic mechanical nonlinearity \cite{Mahboob2012comb,Cao2014,Ganesan2017,Czaplewski2018}; typically, only one or two mechanical modes are utilized to generate low-frequency comb lines through the nonlinear mixing between the driven mode and the parametrically excited harmonic or sub-harmonic modes. In contrast, here, our superconducting electroacoustic system features an array of high-\textit{Q} acoustic modes at 10 GHz frequencies, where the desired Kerr nonlinearity is acquired through the strong electromechanical coupling with a kinetic-inductance dominated superconducting resonator. As a result, cascaded four-wave mixing (FWM) and the subsequent electroacoustic comb generation can be achieved. The hybrid nature of our system also makes the demonstrated electroacoustic comb distinct from the widely studied nanophotonic Kerr microcombs \cite{Kippenberg2011,Jung2016,Yang2017,Kippenberg2018,Suh2016,Yu2018} and microwave microcombs in a long nonlinear superconducting resonator \cite{Erickson2014}, providing a new platform for exploring complex nonlinear phononic-photonic interactions.

\begin{figure*}
\centering
\includegraphics{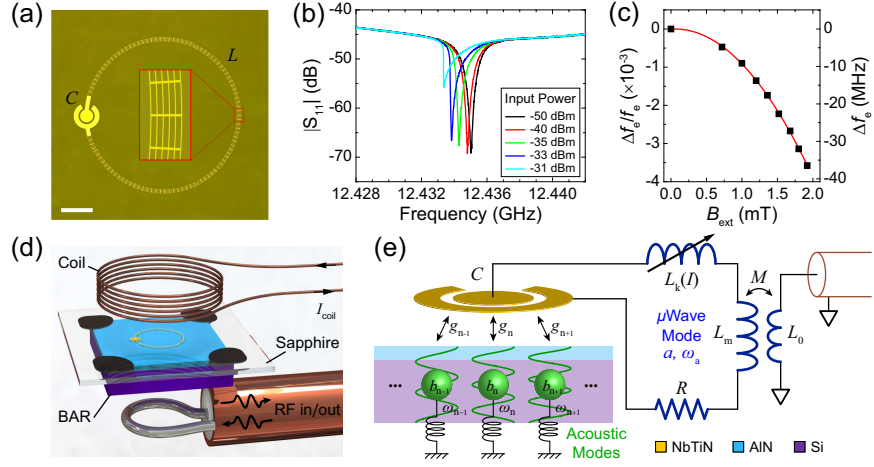}
\caption{(a) An optical image of the nonlinear frequency-tunable Ouroboros resonator. The scale bar is 200\,$\mu$m. Inset: zoomed-in view of the nonlinear inductor (wire width: 1\,$\mu$m, spacing: 4\,$\mu$m). (b) Microwave reflection spectrum of the Ouroboros at different input power showing the nonlinear resonance shift. (c) Frequency tuning of the Ouroboros resonator under an external magnetic field. Red line shows a quadratic fitting. (d) A schematic of the nonlinear electroacoustic device (not to scale). The Ouroboros chip is flipped over and bonded with a piezo-bulk acoustic resonator (BAR) by cryogenic epoxy (black) at the four corners. A superconducting coil is placed above the Ouroboros to provide external magnetic field for frequency tuning. The Ouroboros is inductively coupled to a loop probe for microwave signals input and output. (e) Illustration of the nonlinear electroacoustic system with multimode strong coupling between the Ouroboros and the BAR. $g_n$: piezo-electromechanical coupling strength. $L_\mathrm{k}$ and $L_\mathrm{m}$: the nonlinear kinetic inductance and the magnetic inductance of the Ouroboros. $R$: equivalent resistance (intrinsic dissipation) of the microwave mode.}
\label{fig1}
\end{figure*}

\textit{Nonlinear multimode electroacoustics.--} 
Our nonlinear hybrid system comprises a nonlinear frequency-tunable superconducting Ouroboros resonator \cite{Xu2019} strongly coupled with a bulk acoustic resonator (BAR) through the piezoelectric effect \cite{Han2016,Zou2016}. As the core nonlinear element, the Ouroboros resonator enables two functions: it not only serves as the gain medium for parametric amplification, but also provides frequency tunability to fulfill frequency matching requirement for the comb generation as discussed later. The Ouroboros is fabricated in 50-nm-thick niobium titanium nitride (NbTiN) film on 127-$\mu\textrm{m}$-thick sapphire substrate. As shown in Fig.$\,$\ref{fig1}(a), it is formed by a planar capacitor ($C$) and a nonlinear inductor ($L$) consisting of six narrow wires with high kinetic inductance. The frequency tuning of Ouroboros is achieved by the hole-array structure in the nonlinear inductor; the application of an external magnetic field induces screening DC supercurrents surrounding the holes, which modify the kinetic inductance and hence tune the resonant frequency of the Ouroboros \cite{Xu2019}. As shown in Fig.\,\ref{fig1}(c), a relative frequency tuning of 0.36\% (corresponding to 39 MHz) is realized within 2 mT magnetic field. 

The Kerr nonlinearity of the Ouroboros resonator originates from the quadratic dependence of kinetic inductance $L_\mathrm{k}$ on the current $I$, given by $L_\mathrm{k}(I)=L_\mathrm{k0}\left(1+{I^{2}}/{I_{*}^{2}}\right)$ in the small-current limit \cite{Zmuidzinas2012}. Here, $I_{*}$ is a characteristic current on the order of the critical current of the nanowire. This nonlinearity results in a FWM interaction term in the Hamiltonian described by $H_\mathrm{FWM}=-\hbar\frac{K}{12}(\hat{a}+\hat{a}^{\dagger})^{4}$ with $K=3L_\mathrm{k0}{I_\mathrm{zpf}^{4}}/{I_{*}^{2}}$. Here, $\hat{a}$ ($\hat{a}^\dagger$) is the annihilation (creation) operator of the Ouroboros microwave mode.  $I_\mathrm{ZPF}\equiv\sqrt{{\hbar\omega_\mathrm{e}}/{2L}}$ is the zero point fluctuation current in the nanowire inductor, where $\omega_\mathrm{e}={1}/{\sqrt{LC}}$ is the resonant frequency, and $L=L_\mathrm{k}+L_\mathrm{m}$ with $L_\mathrm{m}$ being the magnetic (or geometric) inductance.

As a manifestation of third-order nonlinearity, the Ouroboros behaves like a Duffing resonator under a strong microwave drive. As shown in Fig.\,\ref{fig1}(b), at low input powers, a symmetric Lorentzian-shape resonance is observed in the reflection spectrum of the Ouroboros. As the input power increases, the resonance starts to distort and shift towards lower frequencies with a bifurcation threshold power around $-33$ dBm. Based on the fitting of the nonlinear resonances \cite{SM}, we extract a Kerr coefficient of ${K}={2\pi}\times 0.11\,\mathrm{mHz}$. It is worth mentioning that, the kinetic inductance nonlinearity also provides opportunity for a three-wave-mixing (TWM) interaction in presence of a DC current. In our experiment, since the pump and the hybrid modes are engineered to only fulfill the FWM condition, the TWM process will not participate.

The BAR consists of a 550-nm-thick aluminum nitride (AlN) layer on top of a 500-$\mu$m-thick high-resistivity silicon substrate. Multiple high-\textit{Q} longitudinal acoustic modes with a free spectral range (FSR) around $\mathcal{F}_0=9.1$\,MHz are supported in the BAR. The thickness of the AlN is chosen to match half acoustic wavelength at 10 GHz to maximize the piezoelectric coupling. To combine the multiple acoustic modes with superconducting nonlinearity, the Ouroboros is flipped over and bonded with the BAR by cryogenic epoxy at the corners (Fig.\,\ref{fig1}(d)). As a result, a strongly coupled nonlinear multimode electroacoustic system is achieved (Fig.\,\ref{fig1}(e)).

\begin{figure*}[ht!]
\centering
\includegraphics{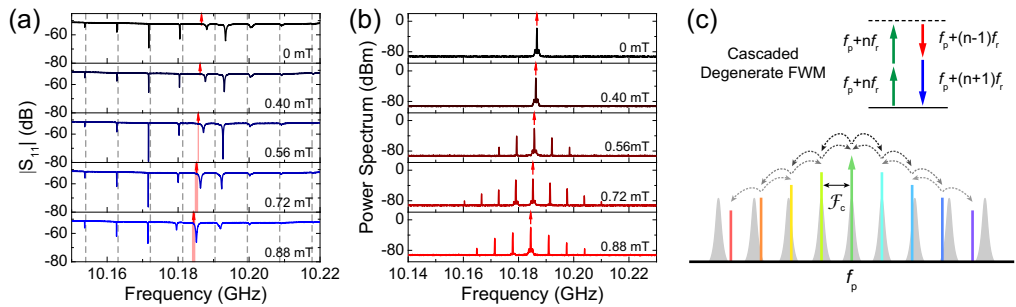}
\caption{(a) Microwave reflection spectrum at different magnetic field. Gray dashed lines indicate the unperturbed acoustic resonant frequencies. Red arrows mark the corresponding pump frequencies in (b). The pink shades indicate the pump frequency range where a hybrid comb can be generated. (b) Microwave power spectrum of the hybrid comb at different magnetic field under a pump power of $-18$ dBm. Red arrows indicated the pump frequency. (c) Principle of comb generation via cascaded degenerate FWM. Under a single pump tone at $f_\mathrm{p}$, two pump photons (green arrows in the upper inset) are annihilated to create one signal and one idler photons (red and blue arrows). In presence of multiple hybrid modes (gray shades), a cascaded process is initiated to form a frequency comb with an FSR of $\mathcal{F}_\mathrm{c}$.}
\label{fig2}
\end{figure*}

The interplay between the Kerr nonlinearity and electro-acoustic coupling can be modeled by the total Hamiltonian of the nonlinear multimode system,
\begin{equation}
\begin{split}
    H &= \hbar\omega_\mathrm{e} \hat{a}^{\dagger}\hat{a} - \frac{\hbar}{2}K\hat{a}^{\dagger}\hat{a}^{\dagger}\hat{a}\hat{a} \\
    &+ \hbar\sum_n \omega_n\hat{b}_n^{\dagger}\hat{b}_n + \hbar\sum_n g_n(\hat{a}^{\dagger}\hat{b}_n+\hat{b}_n^{\dagger}\hat{a}),
\end{split}
\label{Hamil}
\end{equation}
where $\hat{b}_n$ and $\omega_n$ are the annihilation operator and the frequency of the $n$-th acoustic mode, respectively. $g_n$ is the piezo-electromechanical coupling strength between the Ouroboros microwave mode and the $n$-th acoustic mode. Note that we have applied the rotating wave approximation and approximate the FWM interaction by $H_\mathrm{FWM}\approx-\frac{\hbar}{2}K\hat{a}^{\dagger}\hat{a}^{\dagger}\hat{a}\hat{a}$, which appears as the second term in Eq.\,(\ref{Hamil}). It is worth pointing out that the piezo-electromechanical coupling has a linear form, which is distinct from the nonlinear coupling involved in the study of internal resonances in other mechanical systems \cite{Czaplewski2018,Guttinger_2017}.

To better reveal the multimode nonlinear interaction enabled by the strong coupling, it's helpful to describe the system on the hybrid-mode basis. The multimode strong coupling results in hybridization between the microwave mode and the acoustic modes, leading to a set of new hybrid phononic-photonic eigenmodes ($\hat{c}_p$). Mathematically, this corresponds to a unitary transformation which diagonalizes out the linear coupling terms in Eq.\,(\ref{Hamil}) and transforms the Hamiltonian to \cite{SM}
\begin{equation}
    H = \hbar\sum_p {\tilde{\omega}_p\hat{c}_p^{\dagger}\hat{c}_p}
    - \frac{\hbar}{2}\sum_{p,q,r,s}{K_{pqrs}\hat{c}_p^{\dagger}\hat{c}_q^{\dagger}\hat{c}_r\hat{c}_s},
\label{Hamil1}
\end{equation}
Here, $\tilde{\omega}_p$ is the frequency of the \textit{p}-th new hybrid mode. $(p,q,r,s)$ are the dummy summation indices. The effective Kerr coefficient can be expressed as $K_{pqrs}=\frac{K}{\mathcal{N}_p\mathcal{N}_q\mathcal{N}_r\mathcal{N}_s}$, where $\mathcal{N}_p = \sqrt{1+\sum_n{\frac{g_n^2}{(\omega_n-\tilde{\omega}_p)^2+\kappa_{n}^2}}}$ is a normalization factor and $\kappa_n$ is the dissipation rate of the \textit{n}-th acoustic mode (see \cite{SM} for derivation).
Equation\,(\ref{Hamil1}) clearly shows the nonlinear interaction between hybrid modes. As expected, the Kerr nonlinearity is stronger between hybrid modes with more microwave component, i.e. when $\omega_{p,q,r,s}$ is close to $\omega_\mathrm{e}$, because the nonlinearity originates from the superconducting resonator. It is worth pointing out that Eq.\,(\ref{Hamil1}) is valid even in the superstrong coupling regime \cite{Zhang2016,Kostylev2016} where the mode coupling strength becomes comparable with or even larger than the \textrm{FSR}. With this Kerr nonlinearity, a cascaded FWM process can be triggered to generate hybrid phononic-photonic frequency combs.

\begin{figure*}[t!]
\centering
\includegraphics{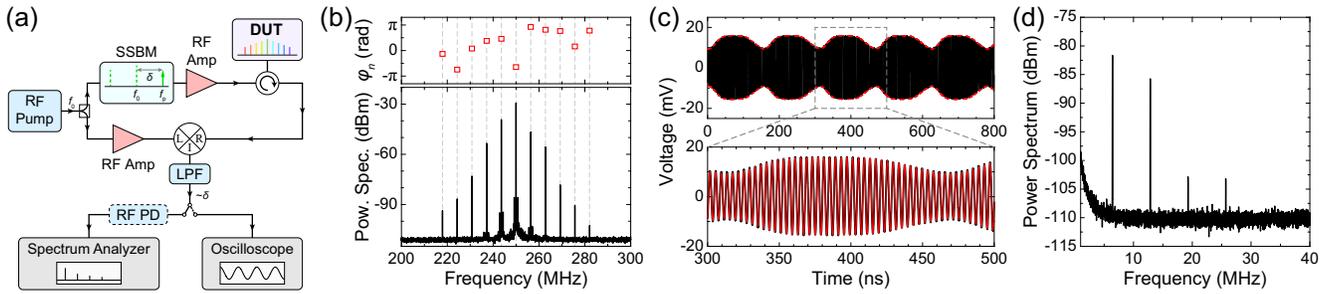}
\caption{(a) Measurement diagram for the comb coherence. An RF pump source ($f_0$\,$\sim$\,10\,GHz) is used to generate a pump tone at $f_\mathrm{p}=f_0+\delta$, where $\delta=250$\,MHz, via single-sideband modulation (SSBM). The 10-GHz frequency comb from the device is then demodulated using the same RF source to down-convert the comb to low frequencies centered at $\delta$. DUT: device under test. RF Amp: RF amplifier. LPF: low-pass filter. PD: photodiode. (b) Microwave power spectrum of the down-converted comb centered at 250 MHz. The top panel shows the relative phase ($\varphi_n$) of each comb line extracted from the fitting of the temporal oscillation in (c). (c) Time-domain oscillation of the down-converted comb. A zoomed-in section with a perfectly matched fitting (red line) is shown in the bottom panel shows. The red dashed lines in the top panel indicate the envelope of entire the fitting curve.  (d) Power spectrum of the mixing signal of the down-converted comb after the RF photodiode.}
\label{fig3}
\end{figure*}

\textit{Hybrid frequency comb.--}
In experiment, the device is enclosed in a copper holder with a superconducting coil placed above the Ouroboros to provide external magnetic field ($B_\mathrm{ext}$) for frequency tuning. Microwave signals are sent into and read out from the Ouroboros by using an inductively coupled loop probe (Fig.\,\ref{fig1}(d)). The device is then mounted on the Still plate of a dilution refrigerator and measured at 900 mK. The passive response of the device is characterized by the microwave reflection spectrum using a vector network analyzer. A low-power probe signal of $-60$ dBm is used to avoid any nonlinear distortion of the spectrum. 

Figure\,\ref{fig2}(a) plots the multimode spectra at different external bias magnetic fields. A set of hybrid modes are clearly observed over a wide frequency range. Due to the dielectric loading of the BAR, the Ouroboros resonant frequency is brought down from $\sim$\,12 to $\sim$\,10 GHz as designed. At zero magnetic field (top panel in Fig.\,\ref{fig2}(a)), the Ouroboros mode at 10.191 GHz is near-resonant with one of the bulk acoustic modes, resulting in normal mode splitting and two most pronounced hybridized modes. The device parameters are extracted by fitting the spectrum using the coupled mode formula (see \cite{SM}). The microwave and the highest acoustic intrinsic \textit{Q} factors are $Q_\mathrm{e,i}\approx3.6\times10^4$ and $Q_\mathrm{m,i}\approx1.6\times10^6$, respectively. The piezo-electromechanical coupling strength is extracted to be $\frac{g_n}{2\pi}\approx2.8$\,MHz, which is larger than the dissipation rates of the microwave and acoustic modes ($g_n>\frac{\kappa_\mathrm{e}}{2},\frac{\kappa_n}{2}$), confirming the multimode strong coupling nature. The FSR of the unperturbed acoustic modes is extracted to be $\mathcal{F}_0=9.1$\,MHz.

Frequency comb generation necessitates a pump tone to provide parametric gain. Since the nonlinearity originates from the superconducting resonator, the pump tone should be applied on a hybrid mode with large microwave component so that sufficient parametric gain can be generated with a reasonable amount of pump power. This means that the hybrid comb generation is most efficient to operate in the strong coupling band where strong hybridization between the microwave mode and the nearby acoustic modes takes place. Nevertheless, strong coupling inevitably perturbs the originally evenly-spaced acoustic resonances \cite{Teufel2011,Han2016,Zhang2016}, leading to non-uniformly distributed hybrid modes with frequency mismatch that hinders FWM and comb generation. This challenge is overcome by the frequency tunability of the Ouroboros, which allows the microwave mode to be adjusted between neighboring bulk acoustic modes to achieve the desired frequency matching condition. 

To investigate the electroacoustic comb generation, a microwave pump of $-18$ dBm is sent to the device. Power spectra of the output signal at different magnetic fields are recorded and shown in Fig.\,\ref{fig2}(b). In order to optimize frequency comb generation, the pump frequency ($f_\mathrm{p}$) is swept across the hybrid mode that has most microwave component (the resonance around the red arrow in Fig.\,\ref{fig2}(a)), which we will refer to as the ``pump mode" hereafter. 
When the pump is tuned into the resonance, the increase of cavity photons will induce resonance shift towards lower frequencies due to the Kerr effect. Therefore, the pump is swept slowly from higher to lower frequencies to make sure the pump can gradually follow and enter the resonance before passing it. 

As shown in Fig.\,\ref{fig2}(b), at zero magnetic field, no comb lines are observed. This is because the asymmetric distribution of the hybrid modes with respect to the pump frequency cannot satisfy the energy conservation requirement for the FWM process---the sum of the signal and idler modal frequencies must equal to twice pump frequency. As the magnetic field increases, the pump mode gradually moves towards lower frequencies. A hybrid comb starts to appear at 0.56 mT and reaches the maximum number of lines at 0.72 mT when the pump mode is tuned to the middle of the two adjacent acoustic resonances.
The FSR of the comb at 0.72 mT is measured to be $\mathcal{F}_\mathrm{c}=6.4$\,MHz, which matches the frequency spacing from the pump to the two nearest neighboring modes in Fig.\,\ref{fig2}(a) but differs from the FSR of the unpereturbed acoustic modes ($\mathcal{F}_0=9.1$\,MHz). This indicates that the comb lines are formed first from the two nearest neighboring modes due to the enhanced density of states at matched frequencies, then expand and get broadened through the cascaded FWM process as illustrated in Fig.\,\ref{fig2}(c). During such process, all the comb lines are generated with well-defined relative phase instead of being random, and hence the comb is expected to be coherent \cite{Herr_2012}.

\textit{Coherence of the comb.--}
The coherence of the electroacoustic comb is first studied by characterizing the temporal oscillation. To measure the time-domain signal, the 10-GHz comb is down-converted to low frequencies centered at 250 MHz. As illustrated in Fig.\,\ref{fig3}(a), the pump tone is generated from a 10-GHz RF source via single-sideband modulation and sent to the device after amplification. The comb signal from the device is demodulated using the same RF source to obtain the down-converted comb (Fig.\,\ref{fig3}(b) lower panel). The low-frequency oscillatory signal is then directly measured by an oscilloscope to obtain the time trace. As shown in Fig.\,\ref{fig3}(c), a stable beating pattern is clearly observed with a beating period measured to be 156 ns, which agrees well with the FSR of the comb ($1/\mathcal{F}_\mathrm{c}\approx156\,\textrm{ns}$). This indicates that each comb line has a stable relative phase ($\varphi_n$); and hence, the hybrid comb is coherent. By fitting the time trace as the sum of sinusoidal oscillations at each comb line frequency, $V=\sum_n A_n \cos(2\pi f_nt+\varphi_n)$ where $A_n$ and $f_n$ are the amplitude and frequency of each comb line, all the relative phases $\varphi_n$ are extracted and plotted in the upper panel in Fig.\,\ref{fig3}(c). More details about the fitting are provided in \cite{SM}.

The coherence of the comb is also revealed by the spectrum of the mixing signal. An RF diode is used to mix all the lines of the down-converted comb. In the case of an incoherent comb, the relative phases between different comb lines will fluctuate randomly and produce noises in the spectrum. In contrast, for a coherent comb, since the comb lines have stable and well-defined relative phases, only sharp peaks at multiple FSRs are expected in the spectrum. Indeed, as shown in Fig.\,\ref{fig3}(d), equally spaced sharp peaks are clearly observed on top of a clean flat background. The frequency of the first peak, as well as the spacing between the adjacent peaks, is measured to be 6.4 MHz, which matches the comb FSR ($\mathcal{F}_\mathrm{c}$), confirming that the electroacoustic comb is coherent.

\textit{Conclusion.--} In summary, we have demonstrated a nonlinear multimode superconducting electroacoustic system at 10 GHz. By harnessing the kinetic inductance nonlinearity and the piezo-electromechanical strong coupling, we realize substantial nonlinear interactions in multimode mechanics at high frequencies and generate a coherent hybrid microcomb. 
The demonstrated nonlinear phononic-photonic platform could open a new route towards the active control of multimode systems as well as advanced information processing in both classical and quantum regimes.

\textbf{Acknowledgment} We acknowledge funding support from Army Research Office (W911NF-18-1-0020). H.X.T. acknowledges support from an NSF EFRI grant (EFMA-1640959) and a Packard Fellowship in Science and Engineering. The authors thank Michael Power and Dr. Michael Rooks for assistance in device fabrication. The authors acknowledge fruitful discussions with Michel Devoret.

\bibliography{XuPhnComb.bib}
\bibliographystyle{XuPRL.bst}


\onecolumngrid
\newpage
\clearpage

\begin{spacing}{1.3}
\begin{center}
  \textbf{\large Supplemental Material for ``Superconducting cavity piezo-electromechanics: the realization of an acoustic frequency comb at  microwave frequencies"}
\end{center}
\end{spacing}

\beginsup

\section{Nonlinearity of the Ouroboros}
The nonlinearity of the Ouroboros can be characterized by measuring the power dependence of the resonance spectrum. As a Kerr resonator, the equation of motion of the Ouroboros mode can be expressed as \cite{Yurke2006}
\begin{equation}
\frac{d\hat{a}}{dt}=-i\omega_\mathrm{e}\hat{a}-\frac{\kappa_\mathrm{e}}{2}\hat{a}+\Big(iK-\frac{\gamma}{2}\Big)\hat{a}^\dagger\hat{a}\hat{a}+\sqrt{\kappa_\mathrm{e,c}}\hat{a}_\mathrm{in}.
\label{eqEOM}
\end{equation}
Here, $\kappa_\mathrm{e}=\kappa_\mathrm{e,i}+\kappa_\mathrm{e,c}$ is the total dissipation rate of the microwave mode, with $\kappa_\mathrm{e,i}$ and $\kappa_\mathrm{e,c}$ being the intrinsic dissipation rate and the external coupling rate, respectively. $\gamma$ is the nonlinear dissipation rate. $\hat{a}_\mathrm{in}$ is the input field, which is related to the output field by $\hat{a}_\mathrm{out}=-\hat{a}_\mathrm{in}+\sqrt{\kappa_\mathrm{ec}}\hat{a}$. Solving Eq.\,(\ref{eqEOM}) in frequency domain, one can express the reflection coefficient as
\begin{equation}
    S_{11,\mathrm{Kerr}}[\omega]=-1+\frac{\kappa_\mathrm{e,c}}{-i(\omega-\omega_\mathrm{e})+\frac{\kappa_\mathrm{e}}{2}
    +(-iK+\frac{\gamma}{2})n},
\label{S11K}
\end{equation}
where $n\equiv\braket{\hat{a}^\dagger\hat{a}}$ is the intra-cavity photon number, which can be solved numerically for a given input field using Eq.\,(\ref{eqEOM}). Equation\,(\ref{S11K}) reveals the nonlinear features of a generic Duffing resonator---as the input power increases, the resonance experiences a frequency shift by $-Kn$, which is proportional to the intra-cavity photon number. 

\begin{figure*}[h!]
\centering
\includegraphics{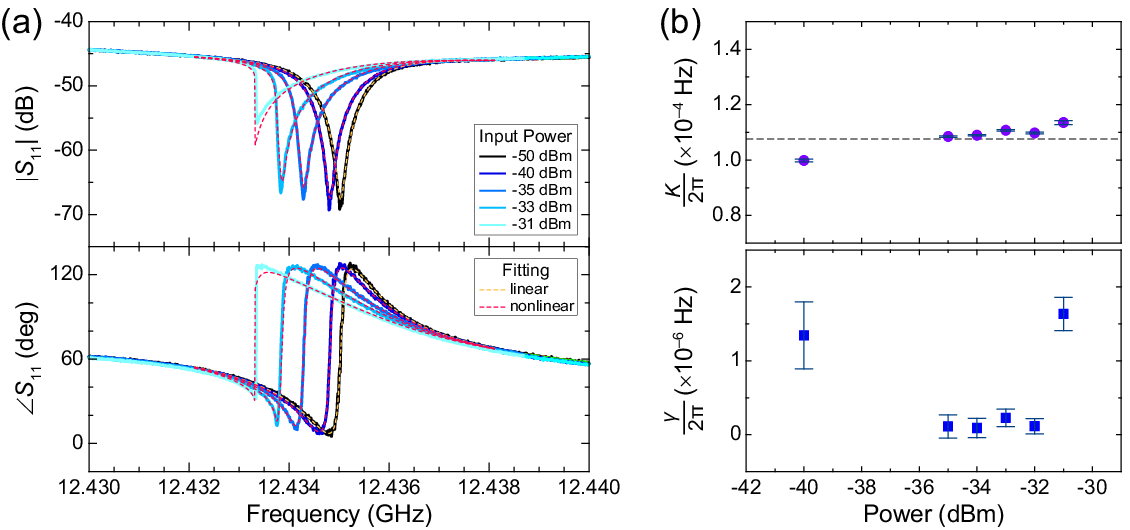}
\caption{(a) Measured Ouroboros reflection spectra at different input powers. The low-power spectrum at $-50$ dBm is fitted as a linear resonator (orange dashed line); whereas the spectra at higher powers are fitted using the Kerr nonlinear formula Eq.\,(\ref{S11K}) (red dashed lines). (b) The extracted Kerr coefficient $K$ and nonlinear dissipation rate $\gamma$ from the fittings at different input powers. The gray dashed line indicates the average value of $\frac{K}{2\pi}=0.11$\,mHz. Fitting error bars are shown on top of data.}
\label{figKerr}
\end{figure*}

To experimentally characterize Kerr nonlinearity, we measure the Ouroboros microwave reflection spectrum at different input power in liquid helium at 4.2 K. Both amplitude and phase responses are plotted in Fig.\,\ref{figKerr}(a). Since the nonlinear distortion of the spectrum can be neglected at low input power, the spectrum at $-50$ dBm is well fitted (orange dashed line) by a linear resonance (i.e., Eq.\,(\ref{S11K}) with $K=\gamma=0$). We therefore extract $\frac{\omega_\mathrm{e}}{2\pi}=12.4351$\,GHz, $\frac{\kappa_\mathrm{e,i}}{2\pi}=0.92$\,MHz, and $\frac{\kappa_\mathrm{e,c}}{2\pi}=0.81$\,MHz. With these values fixed, we fit the nonlinear resonance spectra at higher powers using Eq.\,(\ref{S11K}). As indicated by red dashed lines in Fig.\,\ref{figKerr}(a), the fittings match the experimental data very well---even at the highest power of $-31$ dBm where a jump takes place in the spectrum after the bifurcation threshold. The fitted values of $K$ and $\gamma$ are shown in Fig.\,\ref{figKerr}(b). The average of the Kerr values is taken to obtain $\frac{K}{2\pi}=0.11$\,mHz (labeled by a gray dashed line).



\section{Multimode strong coupling}
\begin{figure*}
\centering
\includegraphics{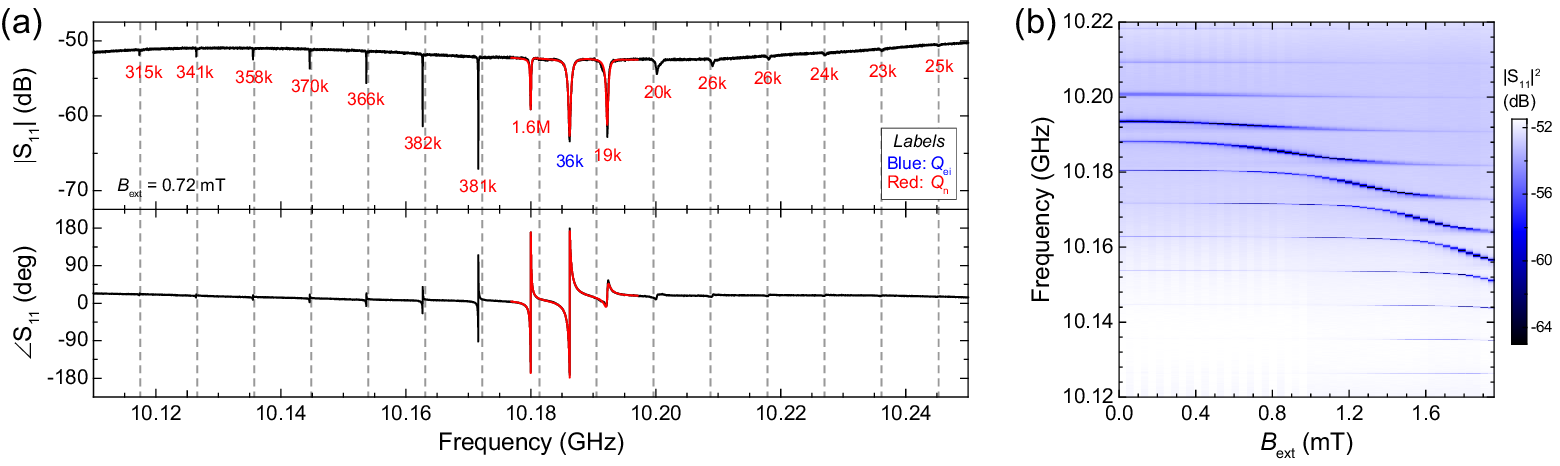}
\caption{(a) Broadband microwave spectrum of the multimode electromechanical system at 0.72\,mT external magnetic field. Red lines show the fitting using the coupled mode formula. The intrinsic microwave and mechanical \textit{Q} factors are labeled in blue and red texts, respectively. (b) 2-D plot of the spectrum at different magnetic field. Avoided crossings are clearly observed when the Ouroboros frequency is tuned over multiple FSRs of the mechanical modes, revealing the signature of multimode strong coupling.}
\label{figMSC}
\end{figure*}

A broadband microwave reflection spectrum of the multimode electroacoustic system is shown in Fig.\,\ref{figMSC}(a) with both amplitude and phase responses. As an example, we plot the spectrum at the external magnetic field $B_\mathrm{ext}=0.72$\,mT when the hybrid comb generation is optimal. The reflection spectrum can be obtained from the coupled-mode theory and given by
\begin{equation}
    S_{11}[\omega]=-1+\frac{\kappa_\mathrm{e,c}}{-i(\omega-\omega_\mathrm{e})+\frac{\kappa_\mathrm{e}}{2}+\sum\limits_n\frac{\abs{g_n}^2}{-i(\omega-\omega_n)+\frac{\kappa_n}{2}}}.
\label{S11}
\end{equation}
Note that since the spectrum is measured with a low input power, the nonlinearity has been neglected.

To extract the device parameters, we fit the section of spectrum containing the three most hybridized modes (i.e. the Ouroboros and two neighbouring acoustic resonances) using Eq.\,(\ref{S11}). As indicated by the red lines in Fig.\,\ref{figMSC}(a), the measured data is fitted with the extracted device parameters listed in Table\,\ref{tab}, where the two acoustic modes are indexed as $n=0$ and 1, respectively.
For far detuned acoustic modes ($\abs{\Delta}\equiv\abs{\omega_n-\omega_\mathrm{e}}\gg\kappa_\mathrm{e},\kappa_n,g_n$), the spectrum can be locally approximated as a Lorentzian line shape
\begin{equation}
    S_{11}[\omega]\Big\rvert_{\omega\sim\omega_n}
    \approx -1+\frac{\frac{\abs{g_n}^2}{\Delta^2}\kappa_\mathrm{e,c}}{-i\big(\omega-\omega_n-\frac{\abs{g_n}^2}{\Delta}\big)
    +\big(\frac{\kappa_n}{2}+\frac{\abs{g_n}^2}{\Delta^2}\frac{\kappa_\mathrm{e}}{2}\big)}.
\label{S11app}
\end{equation}
We use Eq.\,(\ref{S11app}) to fit all other hybrid modes individually. The intrinsic \textit{Q} factors are extracted and labeled in Fig.\,\ref{figMSC}(a) with the highest acoustic \textit{Q} measured to be $Q_\mathrm{m,0}=1.6\times10^6$.

The multimode strong coupling is confirmed by the condition $g_n>(\frac{\kappa_\mathrm{e}}{2},\frac{\kappa_n}{2}$), and the highest cooperativity of $C_n\equiv\frac{4g_n^2}{\kappa_\mathrm{e}\kappa_n}=4.1\times10^3$ is obtained. The multimode strong coupling is further revealed by the two-dimensional (2-D) plot of the spectrum at different magnetic field. As shown in Fig.\,\ref{figMSC}(b), as the external magnetic field increases, the Ouroboros resonance is tuned to lower frequencies over multiple FSRs of the acoustic modes and avoided crossings are observed in sequence as the signature of multimode strong coupling.

\begingroup
\renewcommand{\arraystretch}{1.1} 
\begin{table}[h!]
\begin{center}
\begin{tabular}{||C{14mm}|C{16mm}|C{16mm}|C{16mm}|C{16mm}|C{16mm}|C{15mm}|C{15mm}|C{15mm}|C{15mm}||}
\hline
  &
$\frac{\omega_\mathrm{e}}{2\pi}$\,(GHz)  &  $\frac{\omega_0}{2\pi}$\,(GHz) &
$\frac{\omega_1}{2\pi}$\,(GHz) &
$\frac{\kappa_\mathrm{ei}}{2\pi}$\,(MHz) & $\frac{\kappa_\mathrm{ec}}{2\pi}$\,(MHz) & $\frac{\kappa_0}{2\pi}$\,(MHz) &
$\frac{\kappa_1}{2\pi}$\,(MHz) &
$\frac{g_0}{2\pi}$\,(MHz) &
$\frac{g_1}{2\pi}$\,(MHz) \\  [0.3ex]
\hline\hline
value  &
10.186452  &  10.181356  &  10.190670  &
0.279  &  0.936  &
0.0063  &  0.266  &
2.792  &  2.836  \\
\hline
fit error  &
0.000009  &
0.000004  &
0.000005  &
0.005  &
0.002  &
0.0014  &  0.003  &
0.003  &  0.004  \\
\hline
\end{tabular}
\end{center}
\caption{Fitting results of the spectrum in Fig.\,\ref{figMSC} using the coupled-mode formula Eq.\,(\ref{S11}). The parameters of the Ouroboros microwave mode and two neighbouring acoustic modes are extracted.}
\label{tab}
\end{table}
\endgroup


\section{Effective Kerr coefficient}

In the conventional strong coupling regime, where $g_{n}>\kappa$
but $g_{n}<\mathrm{FSR}$, only the mode $b_{n}$ that is near-resonance
with $a$ needs to be taken into consideration while the influence of other far-detuned modes can be neglected. The resulting hybrid modes are simply the superposition of $a$ and $b_n$ that could be solved analytically.
However, in our experiment, since the coupling strength approaches the FSR of the modes, the system is actually entering into the superstrong coupling regime ($g_{n}\gtrsim\mathrm{FSR}$) where many modes must be considered. The hybrid modes of the system corresponds to the eigenvectors of the high-dimensional matrix and, in general, there is no simple analytical solution. Alternatively, the hybrid modes frequencies could be inferred from the poles of the microwave reflection spectrum given by Eq.\,(\ref{S11}).

Denoting the frequency of the $p$-th new hybrid mode as $\tilde{\omega}_p$, then the hybrid mode itself can be expressed as
\begin{equation}
\hat{c}_p\approx\frac{1}{\mathcal{N}_p}\left[-\sum_n\frac{g_n\hat{b}_n}{-i(\omega_n-\tilde{\omega}_p)+\frac{\kappa_n}{2}}+\hat{a}\right],
\label{hybmode}
\end{equation}
where $\mathcal{N}_p = \sqrt{1+\sum_n{\frac{g_n^2}{(\omega_n-\tilde{\omega}_p)^2+\left(\frac{\kappa_{n}}{2}\right)^2}}}$ is a normalization factor and $\kappa_n$ is the dissipation rate of the \textit{n}-th original acoustic mode.

In analogy to nonlinear optics, the Kerr effect can be understood as a $\chi^{(3)}$ process with the Kerr coefficient determined by the microwave mode overlap in the nonlinear superconducting material. Namely, we have $K=\iiint \chi^{(3)}|\mathbf{u}_a(\mathbf{r}) |^4\mathrm{d}\mathbf{r}$, where $\mathbf{u}_a(\mathbf{r})$ is the microwave mode profile. Based on Eq.\,(\ref{hybmode}), the microwave field in each hybrid mode $\hat{c}_p$ is $\frac{1}{\mathcal{N}_p}\mathbf{u}_a(\mathbf{r})$. Hence, the effective Kerr coefficient among the hybrid modes can be expressed as interaction Hamiltonian can be expressed as
\begin{equation}
K_{pqrs}=\frac{1}{\mathcal{N}_p\mathcal{N}_q\mathcal{N}_r\mathcal{N}_s}\iiint\chi^{(3)}|\mathbf{u}_a(\mathbf{r}) |^4\mathrm{d}\mathbf{r} =\frac{K}{\mathcal{N}_p\mathcal{N}_q\mathcal{N}_r\mathcal{N}_s}.
\end{equation}


\section{Fitting of the temporal oscillation}

The time trace of the comb can be expressed as the sum of sinusoidal oscillations at each comb line frequency $f_n=f_\mathrm{p}^\prime + n \mathcal{F}_\mathrm{c}$, where $f_\mathrm{p}^\prime$ corresponds to the down-converted pump frequency at 250 MHz. Therefore, the time trace (Fig.\,3 in the main text) is fitted by the following equation
\begin{equation}
    V(t) = C\sum_n A_n\cos[2\pi(f_\mathrm{p}^\prime + n \mathcal{F}_\mathrm{c})t + \varphi_n] + B.
\label{Vt}
\end{equation}
Here, $C$ is a global scaling factor and $B$ is a DC constant that is added to account for any imperfect background offset in the measurement. In the fitting, we consider eleven comb lines ($n=-5,...,5$) and fix their amplitudes $A_n$ based on the experimentally measured values (lower panel in Fig.\,3(b) in the main text). All other parameters, including $f_\mathrm{p}^\prime$ and $\mathcal{F}_\mathrm{c}$, in Eq.\,(\ref{Vt}) are set to be free fitting parameters. The fitting results with fitting errors are listed in Table\,\ref{tabFit}. Stable phases of all the eleven comb lines are extracted, confirming the coherence of the electroacoustic comb.

\begingroup
\begin{table}[h!]
\begin{center}
\begin{tabular}{||C{14mm}|C{15mm}|C{15mm}|C{12mm}|C{12mm}|C{12mm}|C{13mm}|C{12mm}|C{12mm}|C{12mm}|C{12mm}|C{12mm}||}
\hline
  &
$\varphi_{-5}$  &
$\varphi_{-4}$  &
$\varphi_{-3}$  &
$\varphi_{-2}$  &
$\varphi_{-1}$  &
$\varphi_{0}$  &
$\varphi_{1}$  &
$\varphi_{2}$  &
$\varphi_{3}$  &
$\varphi_{4}$  &
$\varphi_{5}$  \\ 
\hline\hline
value  &
$-0.4$  &  $-2.4$  &  2.44  &  1.186  &
1.438  &  $-2.0423$  &  2.896  &  2.552  &
2.40  &  5.0  &  2.4  \\
\hline
fit error  &
0.3  &  0.1  &  0.03  &  0.003  &
0.001  &  0.0003  &  0.002  &  0.004  &
0.05  &  0.2  &  0.8  \\ 
\hline \noalign{\vskip 3mm} \hline
  &
$f_\mathrm{p}^\prime$\,(MHz)  &
$\mathcal{F}_\mathrm{c}$\,(MHz)  &
$B$\,(mV)  &  $C$  &
  &  &  &  &  &  &  \\ 
\hline\hline
value  &
250.0009  &  6.4139  &  0.225  &  0.37674  &
  &  &  &  &  &  &  \\
\hline
fit error  &
0.0001  &  0.0003  &  0.002  &  0.00006  &
  &  &  &  &  &  &  \\ 
\hline
\end{tabular}
\end{center}
\caption{Fitting results of the oscillatory time trace using Eq.\,(\ref{Vt})}
\label{tabFit}
\end{table}
\endgroup

\end{document}